\documentclass[reprint,superscriptaddress,amsmath,amssymb,
aps,prb]{revtex4-2}
\usepackage{hyperref}
\hypersetup{colorlinks,allcolors=blue}
\usepackage{dcolumn}
\usepackage{bm}
\usepackage{tabularx}
\newcolumntype{Y}{>{\centering\arraybackslash}X}
\usepackage{wrapfig}
\usepackage{graphicx} 
 \usepackage{bm}

\graphicspath{{pictures/}}

\begin{document}
\title{Domain wall dynamics of ferrimagnets induced by spin-current near the angular momentum compensation temperature}
\author{V.V.~Yurlov}
\affiliation{Moscow Institute of Physics and Technology, Institutskiy per. 9, 141700 Dolgoprudny, Russia} 
\author{K.A.~Zvezdin}
\email{zvezdin.ka@phystech.edu}
\affiliation{Prokhorov General Physics Institute of the Russian Academy of Sciences, Vavilova 38, 119991 Moscow, Russia}
\affiliation{New Spintronic Technologies, Russian Quantum Center, Bolshoy Bulvar 30, bld. 1, 121205  Moscow, Russia}
\author{P.N.~Skirdkov}
\affiliation{Prokhorov General Physics Institute of the Russian Academy of Sciences, Vavilova 38, 119991 Moscow, Russia}
\affiliation{New Spintronic Technologies, Russian Quantum Center, Bolshoy Bulvar 30, bld. 1, 121205  Moscow, Russia}
\author{A.K.~Zvezdin}
\affiliation{Prokhorov General Physics Institute of the Russian Academy of Sciences, Vavilova 38, 119991 Moscow, Russia}

\bibliographystyle{apsrev4-2}
\date{\today}

\begin{abstract}
We report on a theoretical study of the spin-current excited dynamics of domain walls (DWs) in ferrimagnets in the vicinity of the angular momentum compensation point. Effective Lagrangian and nonlinear dynamic equations are derived for a two-sublattice ferrimagnet taking into account both spin-torques and external magnetic field. The dynamics of the DW before and after the Walker breakdown is calculated for any direction of the spin current polarization. It is shown that for the in-plane polarization of the spin current, the DW mobility reaches a maximum near the temperature of the angular momentum compensation. For the out-of-plane spin polarization, in contrast, a spin current with the densities below the Walker breakdown does not excite the dynamics of the DW. After overcoming the Walker breakdown, the domain wall velocity increases linearly with increasing the current density. In this spin-current polarization configuration the possibility of a gigahertz oscillation dynamics of the quasi-antiferromagnetic vector under the action of a damping-like torque in the angular momentum compensation point is demonstrated. Possible structures for experimental demonstration of the considered effects are discussed.
\end{abstract}

\maketitle
\section{Introduction}

Spintronics, which is a rapidly developing branch of nanoelectronics, is based on the concept that the principal role in information processing belongs to spins of electrons instead of charges\cite{doi:10.1063/1.1578165,  doi:10.1063/1.1594841}. The mainstream of spintronics is attaining the winning combination of the spin transport efficiency and nanoscale size of spintronic devices. In this regard, magnetic DWs attract increasing attention\cite{ c1, PhysRevB.70.024417, doi:10.1063/1.3671438}: they can be used to store and transmit information in race track or magnetic random access memories (MRAM)\cite{Parkin190, doi:10.1063/1.5078525, PhysRevLett.102.067206, doi:10.1063/1.4883740, chanthbouala2011vertical}. 
\par
Conventional spintronic devices use ferromagnetic (FM) materials owing to their property to create and subsequently to use spin-polarization of the conducting electrons. The nanural restrictions of ferromagnetic spintronic devices relate to the limited operation frequencies and general energy efficiency.  Recent advances in spin current injection into insulating antiferromagnets (AFMs) have revealed the prospects of AFM spintronics, whose advantages is an extremely high frequency in comparison to the operation frequency of ferromagetic devices\cite{Wadley587, RevModPhys.90.015005}. At the same time the AFM spintronics has its own shortcomings associated with the difficulties to detect the magnetization states and magnetization dynamics. These motivate a boosting development of  ferrimagnetic (FiM) spintronics, which combines the ultra-high operation frequences close to those of AFM devices, with much more reliable ways to detect its magnetization states. Very rich and interesting magnetization dynamics\cite{PhysRevB.74.134404, PhysRevB.73.220402}, in terms of fundamental and applied physics, is observed in these materials near the points of compensation of magnetization and angular momentum. Moreover, by manipulating the temperature of the ferrimagnet near the compensation points, outstanding magnetization switching characteristics can be obtained\cite{doi:10.1063/5.0010687, PhysRevB.100.064409, PhysRevApplied.13.034053}. It has been shown that the electrical current can be an efficient approach to magnetization switching\cite{PhysRevLett.119.077702, doi:10.1063/1.4962812, PhysRevLett.118.167201, PhysRevApplied.6.054001}. GdFeCo FiM layer demonstrates ultrafast magnetization reversal influenced by femtosecond laser pulses in various experiments\cite{Bokor}. These results suggests that the FiMs based structures can form a promising technological platform for ultrafast spintronic memory devices. 

Angular momentum compensation point $T_A$, where $M_1/\gamma_1 = M_2/\gamma_2$, $\gamma_i$ is the gyromagnetic ratio of the i-sublattice (i = 1, 2), represents a very promising line of research of FiMs magnetization dynamics\cite{PhysRevB.86.214416, PhysRevB.74.134404, 4671127}. Recent field-driven experiments demonstrated high velocity and great mobility of a domain wall in FiM near the $T_A$\cite{Kim2017}. The next natural step in this direction is to use the spin-currents to manipulate DWs position and dynamics\cite{PhysRevLett.121.057701}. While the spin-current induced phenomena in FMs seems to be well comprehensible, the mechanisms of spin transfer in AFMs and compensated FiMs are still not figured out properly. 

In the present research we develop a model to describe DW motion in ferrimagnets near the angular momentum compensation point in case of arbitrary spin current polarization and torque type. DW dynamics influenced by a spin-current in FiMs is studied generally by using collective coordinates model and Landau-Lifshitz-Gilbert equation with addition of spin-transfer torque components\cite{PhysRevB.96.100407, MARTINEZ2019165545, doi:10.1063/10.0001552}. Here instead we employ the Lagrangian formalism for two subblatice ferrimagnet. This approach allows us to strictly define ferrimagnetic parameters such as width of the DW, velocity of the magnons, transverse magnetic susceptibility, effective Gilbert damping parameter and gyromagnetic ratio by using perturbation theory. 
\par
We derive non-linear dynamic equations based on the Lagrangian formalism, which is similar to Slonszewski equations\cite{Slon}. Using this model, we calculate the dynamics of the domain wall in FiM, depending on the direction of the polarizer, electric current density and temperature, before and after the Walker breakdown. In our modeling we observe that for the in-plane polarization of the spin current, the DW mobility reaches a maximum near the temperature of the angular momentum compensation, and vanishes after bypassing the Walker breakdown. For the out-of-plane spin polarization, in contrast, a spin-current with the densities below the Walker breakdown does not excite the stationary dynamics of the DW. After overcoming the Walker breakdown, the domain wall velocity increases linearly with increasing the electric current density. In this configuration of the spin current, near the compensation point $T_A$ we observe gigahertz oscillations of the quasi-antiferromagnetic vector. 

\section{Model and basic equations}
\begin{figure}[h!]
\begin{center}
\center{\includegraphics[width=0.9\linewidth]{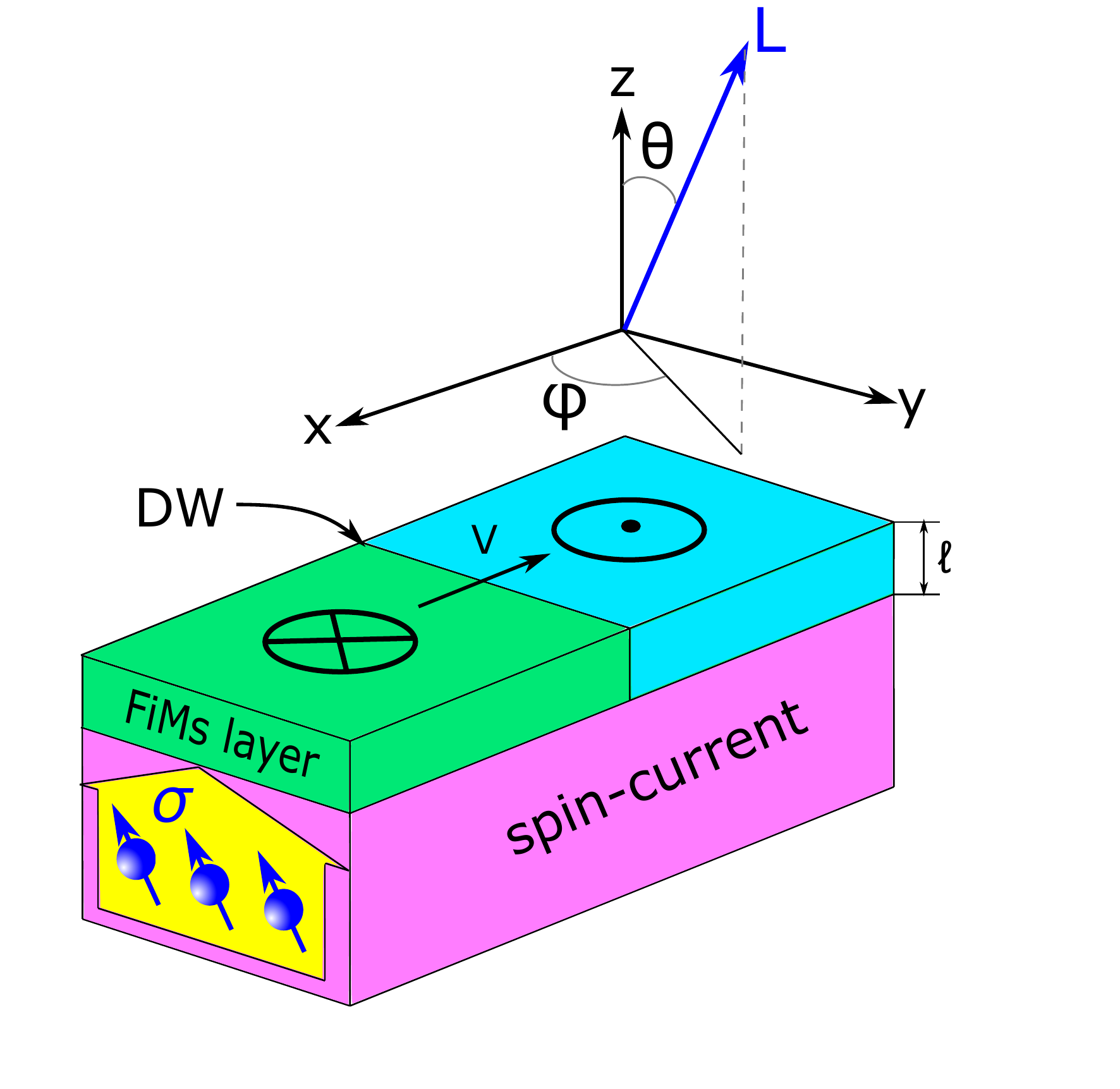}}
\caption{Schematic of considered FiMs with single domain wall, $\bm{\sigma}$  is polarization vector of the spin-current, $l$ - is thickness of the sample; $\theta$ and $\varphi$ are the polar and azimuthal angels of quasi-antiferromagnetic vector $\textbf{L}$.}
\label{Fig1}
\end{center}
\end{figure}
We develop a model based on the Lagrange formalism for describing DW dynamics due to spin-current. For two sublatticed FiMs ordering parameters related to magnetizations $\textbf{M}_1$, $\textbf{M}_2$ of these sublattices can be introduced in the vicinity of compensation temperatures as quasi-antiferromagnetic vector $\textbf{L} = \textbf{M}_1 - \textbf{M}_2$ and ferromagnetic $\textbf{M} = \textbf{M}_1 + \textbf{M}_2$  order parameters. We consider a spin-current with a polarization $\bm{\sigma}$ flowing through the FiM film (see Fig.~\ref{Fig1}). By analogy with the approach used for ferromagnets, we use the adiabatic approximation, assuming that the AFM  order parameter $L$ changes slowly in comparison with the spins of the injected electrons. The spin-current excites a spin transfer torque acting on local magnetization of the i-sublatice respectively and in general case composed to the in–plane and the out–of plane components $\textbf{T}^{i}_{ST} = \textbf{T}^{i}_{FL} + \textbf{T}^{i}_{DL}$. The $\textbf{T}^{i}_{FL}$ component due to its symmetry is usually refereed as a field–like torque and has the following form: $\textbf{T}^{i}_{DL} \sim [\textbf{M}_i \times \bm{\sigma}]$. The $\textbf{T}^{i}_{DL}$ torque component has a symmetry similar to the damping torque and usually refereed as an damping-like torque (or anti-damping-like torque): $\textbf{T}^{i}_{DL} \sim [\textbf{M}_i\times[\textbf{M}_i \times \bm{\sigma}]]$. Typically the magnitude of the anti-damping-like torque component is significantly lager than the field-like one for magnetic tunnel junctions, however in case of spin-orbit torques can be of a similar magnitude. 

The magnetization dynamics is described by a system of Euler-Lagrange equations:
\begin{equation}\label{eq1}
    \begin{cases}
    \begin{gathered}
   \dfrac{d}{dt}\Big(\dfrac{\partial\mathcal{L}}{\partial\dot{\theta}_i}\Big) - \dfrac{\delta\mathcal{L}}{\delta\theta_i} = -\dfrac{\partial \mathcal{R}_{i}}{\partial \dot{\theta}_i} - \dfrac{\partial W}{\partial\dot{\theta}_i}\\
   \dfrac{d}{dt}\Big(\dfrac{\partial\mathcal{L}}{\partial\dot{\varphi}_i}\Big) - \dfrac{\delta\mathcal{L}}{\delta\varphi_i} = -\dfrac{\partial \mathcal{R}_{i}}{\partial \dot{\varphi}_i} - \dfrac{\partial W}{\partial\dot{\varphi}_i}
   \end{gathered}
   \end{cases},
\end{equation} 
where $\mathcal{L}$ and $\mathcal{R}_i$ are the Lagrangian and Rayleigh functions, $W$ is spin transfer torque power density; $\theta_i$ and $\varphi_i$ are the polar and azimuthal angles characterizing the orientation of the i-th sublattice magnetization (i = 1, 2). Note that $\delta W$ represents external spin-current effect on magnetic structure and consists of the damping-like and the field-like components. Due to its symmetry the field-like component can be included in the Lagrangian by using the quasi-antiferromagnetic approximation. Thus, $\delta W$ in the Euler-Langrange equations consist of only damping-like spin-current component. Hereinafter we turn to the effective Lagrangian $\mathcal{L}_{\mathrm{eff}}$, effective Rayleigh function $\mathcal{R}_{\mathrm{eff}}$ and power density of a spin current $\delta W$ in quasi-antiferromagnetic approximation applicable in the vicinity of compensation temperatures (see in Supplementary)\cite{Davydova_2019} 
\begin{equation}\label{eq2}
    \begin{gathered}
   \mathcal{L}_{\mathrm{eff}} = \dfrac{\chi_{\perp}}{2}\Big(\dfrac{\dot{\theta}}{\overline{\gamma}_{\mathrm{eff}}}\Big)^2 + m\Big(H - \dfrac{\dot{\varphi}}{\gamma_{\mathrm{eff}}}\Big)\cos\theta  + \\ \dfrac{\chi_{\perp}}{2}\Big(H - \dfrac{\dot{\varphi}}{\overline{\gamma}_{\mathrm{eff}}}\Big)^2\sin^2\theta - K_u\sin^2\theta-\\ - K_{\perp}\sin^2\theta\sin^2\varphi -  A\Big(\Big(\dfrac{d\theta}{dx}\Big)^2 + \sin^2\theta\Big(\dfrac{d\varphi}{dx}\Big)^2\Big) - \\ - \dfrac{\chi_{\perp}}{2}\Big(\dfrac{\overline{B}}{\mathcal{M}}\Big)^2\Big(\sin^2\theta n_{\perp} + \\ + \cos^2\theta\cos^2(\varphi -\psi)n_{\parallel} + \sin^2(\varphi - \psi)n_{\parallel}\Big),\\
   R_{\mathrm{eff}} = \dfrac{\alpha_{\mathrm{eff}} \mathcal{M}}{\gamma_{\mathrm{eff}}}\Big(\dot{\theta}^2 + \sin^2\theta\cdot\dot{\varphi}^2\Big),\\
   \delta W = -\overline{A}\sin(\varphi - \psi)n_{\parallel}\cdot\delta\dot{\theta} + \\ + (-\overline{A} n_{\perp}\sin^2\theta + \overline{A} n_{\parallel}\cos\theta\cos(\varphi - \theta))\cdot\delta\dot{\varphi},
   \end{gathered}
\end{equation}
where $m = M_2 - M_1$, $\mathcal{M} = M_1 + M_2$; $\chi_{\perp} = \mathcal{M}/H_{ex}$ is transverse magnetic susceptibility, $H_{ex}$ is an exchange magnetic field acting between sublattices; $K_u$ and $K_{\perp}$ are constants of uniaxial and in-plane magnetic anisotropies respectively; A is an exchange stiffness constant, $\theta$ and $\varphi$ are the polar and azimuthal angles of an quasi-antiferromagnetic vector $\textbf{L}$, $\textbf{H} = (0, 0 , H_z)$ is a magnetic field applied along the “easy magnetization axis”; $\alpha_{\mathrm{eff}} = \overline{\alpha}m/(m - m_0) $, $\gamma_{\mathrm{eff}} = \overline{\gamma}m/(m - m_0)$, $\overline{\gamma}_{\mathrm{eff}} = \overline{\gamma}\cdot(1 - m\cdot m_0/\mathcal{M}^2)^{-1}$, $\overline{\alpha} =(\alpha_1\gamma_2 + \alpha_2\gamma_1)/2(\gamma_1 + \gamma_2)$, $1/\overline{\gamma} = (1/\gamma_1 + 1/\gamma_2)/2$, where $\alpha_i$ and $\gamma_i$ are a damping constant and a gyromagnetic ratio for the i-sublatice respectively, $m_0 = \mathcal{M}(\gamma_1 - \gamma_2)/(\gamma_1 + \gamma_2)$; $\overline{A} = \hbar J P_{DL}/(2 e l)$ and $\overline{B} = \hbar J P_{FL}/(2 e l)$ are the field-like and the damping (or anti-damping) spin transfer torque coefficients, where $J$ is electrical current density, $l$ is the thickness of the magnetic film, $e > 0$ is the electron charge; $n_{\perp}$ and $n_{\parallel}$ are the out-of- and the in- plane components of unit vector $\textbf{n} = (n_x, n_y, n_z)$ along the polarization of spin-current $\bm{\sigma}$, $\psi$ is an angle between the projection of  polarization vector of the spin-current  $\bm{\sigma}$ on the x-y plane and the x-axis; $P_{DL}$ and $P_{FL}$ are the field-like and the damping (or anti-damping) polarizations of the spin current, respectively.
\par
Implementing the procedure which is described in the Supplementary, we derive the system of dynamic equations for the $\mathrm{180^{\circ}}$ DW without external magnetic field:
\begin{equation}\label{eq3}
    \begin{cases}
    \dfrac{2\overline{\alpha} \mathcal{M}}{\overline{\gamma}\Delta_0}\dot{q} + m\dfrac{\dot{\varphi}}{\gamma_{\mathrm{eff}}}= \widetilde{T_{\theta}} \\
    -\dfrac{\chi_{\perp}}{\overline{\gamma}^2_{\mathrm{eff}}}\ddot{\varphi} + \dfrac{m}{\gamma_{\mathrm{eff}}}\dfrac{\dot{q}}{\Delta_0} - K_{\perp}\sin2\varphi -  \dfrac{2\overline{\alpha} \mathcal{M}}{\overline{\gamma}}\dot{\varphi} = \widetilde{T_{\varphi}}
    \end{cases},
\end{equation}
 where q is a coordinate of the DW centre, $\Delta_0 = \sqrt{A/K_u}$ is a width of the DW. The spin transfer torque components are written as
 \begin{equation}\label{eq4}
    \begin{gathered}
     \widetilde{T_{\theta}} = -\dfrac{\pi}{2}\overline{A}\sin(\varphi - \psi)n_{\parallel}\\
     \widetilde{T_{\varphi}} = -\overline{A} n_{\perp}  + \dfrac{\chi_{\perp}}{2}\Big(\dfrac{\overline{B}}{\mathcal{M}}\Big)^2\sin2(\varphi - \psi)n_{\parallel}
     \end{gathered}.
 \end{equation}
  Note that in general case the width of the DW is determined as $\Delta = \Delta_0 \sqrt{1 - (\dot{q}/c)^2}$, where $c =\overline{\gamma}_{\mathrm{eff}} \sqrt{2A/\chi_{\perp}}$ is a magnons velocity (see in Supplementary). For our set of parameters it can be estimated as $c \sim 8$ km/s. As a result, the variation of the DW width for the considered velocities is of the order of one percent ($\Delta /\Delta_0 \sim 0.01$). Thus we can assume that $\dot{q} \ll c$ and DW width $\Delta \approx \Delta_0$. 

\section{Dynamic Equation Analysis}

To understand peculiar features of the current induced DW dynamics in compensated FiMs following from eqs. (\ref{eq3}) and (\ref{eq4}) we analyze several particular cases. To calculate the DW dynamics we use typical GdFeCo parameters: \cite{Kim2017}:
   $ K_u \sim 1\cdot10^5$ erg/cc, $\mathcal{M} \approx 900$ emu/cc,  $\overline{\alpha} \sim 0.02$, $\overline{\gamma} \sim 2 \cdot 10^7$, $A \sim 1 \cdot 10^{-6}$ erg/cm, $g_d = 2.2$, $g_f = 2$, $T_M = 220$ K, $T_A = 310$ K, $l = 10$ nm, where $g_d$ and $g_f$ are Lande g-factors for d- and f-sublatices respectively. The constant of in-plane magnetic anisotropy in case of infinite film is $K_{\perp} = 2 \pi m^2$, however in case of a narrow FiMs nanowire it has a different form due to magnetostatic interaction. Note, that all dynamic parameters (such as velocity, DW displacement and others) are functions of $\nu = m/\mathcal{M}$, which can be rewritten in term of temperature $T$ by using the following expression :
\begin{equation}\label{eq5}
    \nu = \dfrac{m}{\mathcal{M}} = \dfrac{T - T_M}{T^*},
\end{equation}
where $T^* = 1891$ K is obtained from the GdFeCo parameters\cite{Kim2017}. For all further mentioned modelling results $P_{DL}=0.3$ and $P_{FL}=0.03$. 
\par
\begin{figure}[h!]
\begin{center}
\center{\includegraphics[width=1\linewidth]{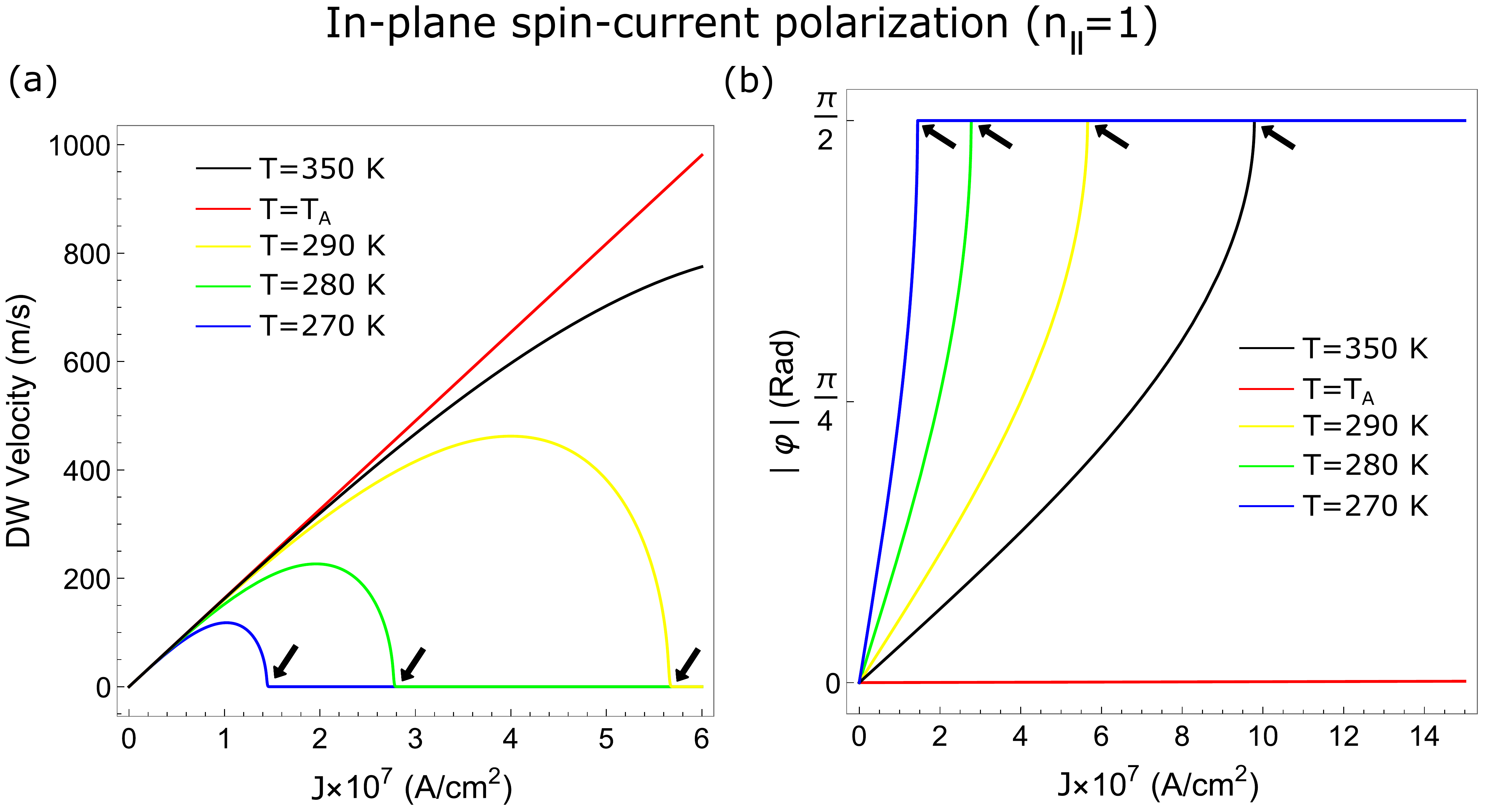}}
\caption{a) Average DW velocity in Walker and post Walker regimes as a function of the electrical current density J; b) Absolute value of the azimuthal angle $\varphi$ in Walker and post Walker regimes as a function of the electrical current density J; blue, green, yellow, red and black curves correspond to the temperatures $T = 270$ K, $T = 280$ K, $T = 290$ K, $T = T_A$ and $T = 350$ K respectively; the black arrow indicate the transition in the post Walker regime. All curves are plotted for the in-plane spin current polarization.}
\label{Fig5}
\end{center}
\end{figure}
\par
First, let us analyze DW dynamics for the in-plane spin-current when $K_{\perp} \neq  0$ and the spin polarization along the y axis $n = (0, 1, 0)$ ($\psi = \pi/2$ and $n_{\parallel} = 1$). In this geometry stationary DW motion ($\dot{\varphi} = 0$ and a constant DW) is observed below Walker breakdown. In the $T_A$ the azimuthal angle $\varphi$ tends to zero (see red curve in Fig.~\ref{Fig5}(b)) and the stationary DW motion is observed with the velocity $\dot{q}/\Delta_0 = \pi\overline{\gamma}\overline{A}/4\overline{\alpha}\mathcal{M}$, which follows from the first equation in (\ref{eq3}). As follows from (\ref{eq3}) and (\ref{eq4}) in this case the damping-like (or anti-damping-like) spin transfer torque component with magnitude $\overline{A}$ initiates DW dynamics, while the field-like one  only modifies the magnetostatic term. The magnitude of the azimuthal angle $\varphi$ increases with increasing in the electric current density and tends to $\pi/2$ which is demonstrated in Fig.~\ref{Fig5}(b). 
Note that in the case of the in-plane polarizer after reaching the critical current density corrisponding to the Walker breakthrough $J^* = 16 e l |\alpha_{\mathrm{eff}}| K_{\perp}/\pi \nu \hbar P_{AD}$, there is no domain wall motion observed - it is indicated by the black arrows in the Fig.~\ref{Fig5}(a) and Fig.~\ref{Fig5}(b). This range corresponds to the constant azimuthal angle $\varphi \approx \pi/2$. Spin-current cannot push the domain-wall when angle $\varphi$ exceeds $\pi/2$ (see Fig.~\ref{Fig5}(b)) for the case of in-plane polarization. This means that the steady precessional motion of DW is impossible for in-plane polarized spin current and DW velocity eventually drops to zero
for all temperatures except for $T_A$.
\par
Now let us discuss a more difficult situation when the $\bm{\sigma}$ is parallel to the z-axis $\textbf{n} = (0, 0, 1)$ and $n_{\perp} = 1$. Actually, if we assume that $K_{\perp} = 0$, $\chi_{\perp} \ll 1$ and consider the temperatures in the vicinity of angular momentum compensation point $T_A$, the system (\ref{eq4}) describes the steady motion of the DW with velocity $\dot{q}$ and precession rate $\dot{\varphi}$:
\begin{equation}\label{eq6}
    \begin{gathered}
    \dfrac{\dot{q}}{\Delta_0} =  -\dfrac{\overline{\gamma}}{2\mathcal{M}\overline{\alpha}}\cdot\dfrac{ \overline{A}\nu/2\alpha_{\mathrm{eff}}}{1 +(\nu/2\alpha_{\mathrm{eff}})^2},
    \end{gathered}
\end{equation}
\begin{equation}\label{eq7}
    \begin{gathered}
    \dot{\varphi} = \dfrac{\overline{\gamma}}{2\mathcal{M}\overline{\alpha}}\cdot\dfrac{ \overline{A}}{1 +(\nu/2\alpha_{\mathrm{eff}})^2}.
    \end{gathered}
\end{equation}
\par
\begin{figure}[ht]
\center{\includegraphics*[width=1\linewidth]{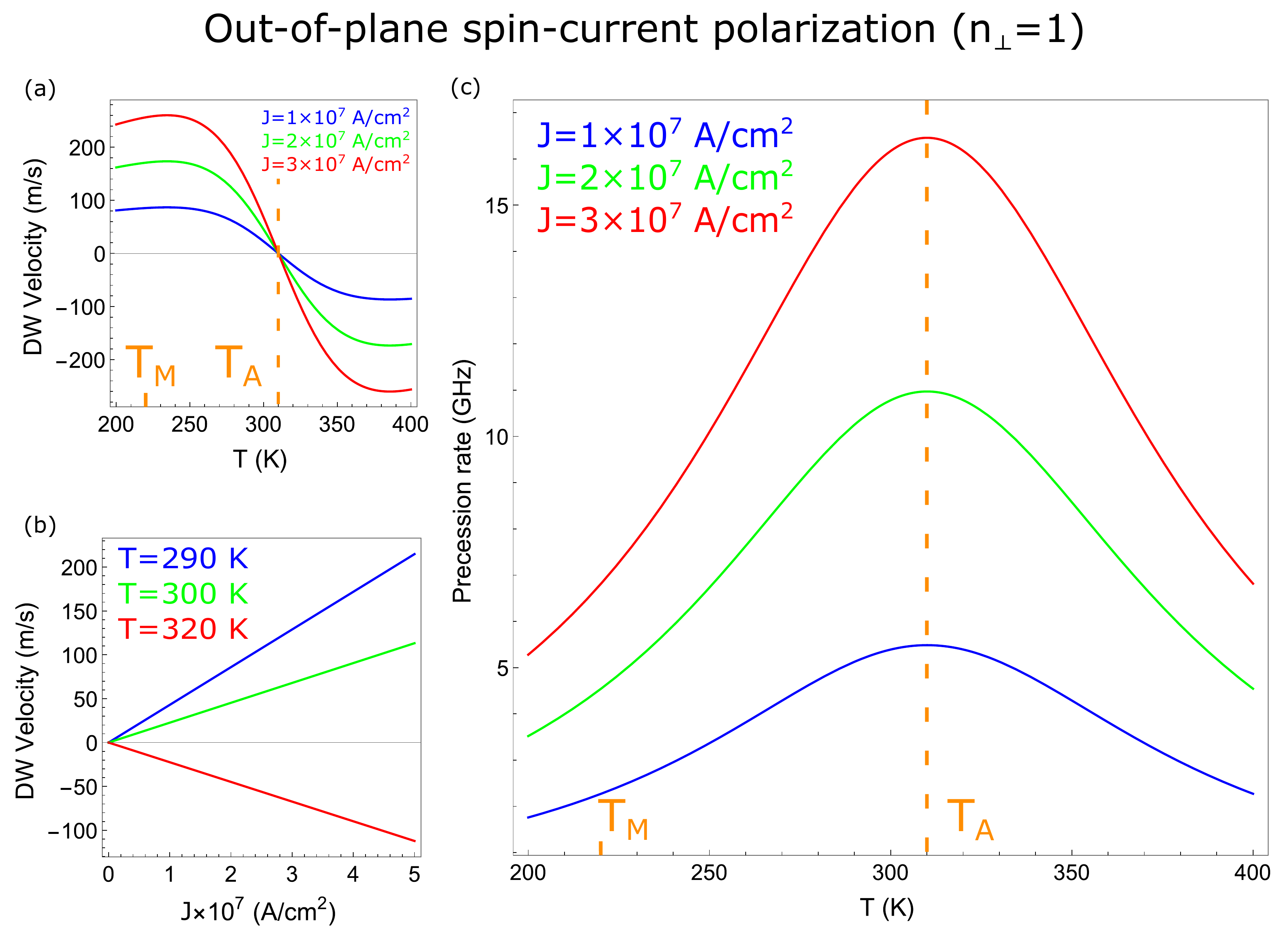}}
\caption{a) Dependence of the DW velocity on temperature at the different current densities; b) Dependence of the DW velocity on electrical current density at the different temperatures; c) Precession rate as a function of temperature at the different current densities. All curves are plotted for the out-of-plane spin current polarization.}
\label{Fig2}
\end{figure}

By using the equation (\ref{eq5}) we can rewrite the (\ref{eq6}) and (\ref{eq7}) in term of temperature T and study the dependence of the DW velocity and precession rate on temperature and current density. Fig.~\ref{Fig2}(a) demonstrates that the DW velocity has two  maximum values near the angular momentum compensation point and these values increase with growth in current density. These curves are asymmetric with respect to $T_A$. Thus, the velocity of the DW changes its sign passing through the angular momentum compensation point. This situation is also realized in Fig.~\ref{Fig2}(b), where the dependence of the DW velocity on electric current density at different temperatures is given. As it is seen from the equation (\ref{eq6}) DW velosity linearly depends on the electrical current density. The blue and green line (see Fig.~\ref{Fig2}(b)) lie below the $T_A$ and the slope of this curves (DW mobility $\dot{q}/J$) decreases. The DW velocity changes its direction above the angular momentum compensation point (red curve in Fig.~\ref{Fig2}(b)).
\par
Note, that the DW velocity reaches 260 m/s at current densities by about $3\times10^7$ A/cm$^2$. Precession rate is not zero $\dot{\varphi} \neq 0$ in the vicinity of the angular momentum compensation temperature compared with field-driving DW motion\cite{ZVEZDIN2020166876} (where magnetic field is applied along the easy magnetization axis).  The equation (\ref{eq7}) shows that $\dot{\varphi}$ reaches its maximum by about 17 GHz at low current density ($\sim 3\times10^7$ A/cm$^2$) near the  $T_A$ (see in Fig.~\ref{Fig2}(c)). As it follows from the equations (\ref{eq6}) and (\ref{eq7}) in case of considered polarization direction both oscillation of the $\varphi$ angel and DW motion is triggered by the damping (or anti-damping) spin transfer torque component with magnitude $\overline{A}$; hence DW dynamic and oscillation freezes without spin-current. Field-like spin transfer torque is neglected at the out-of-plane spin-current polarization case due to decomposition of Lagrangian of the two-sublattice ferrimagnet as a next order small parameter (see in Supplementary). 
\par
\begin{figure}[h!]
\begin{center}
\center{\includegraphics[width=1\linewidth]{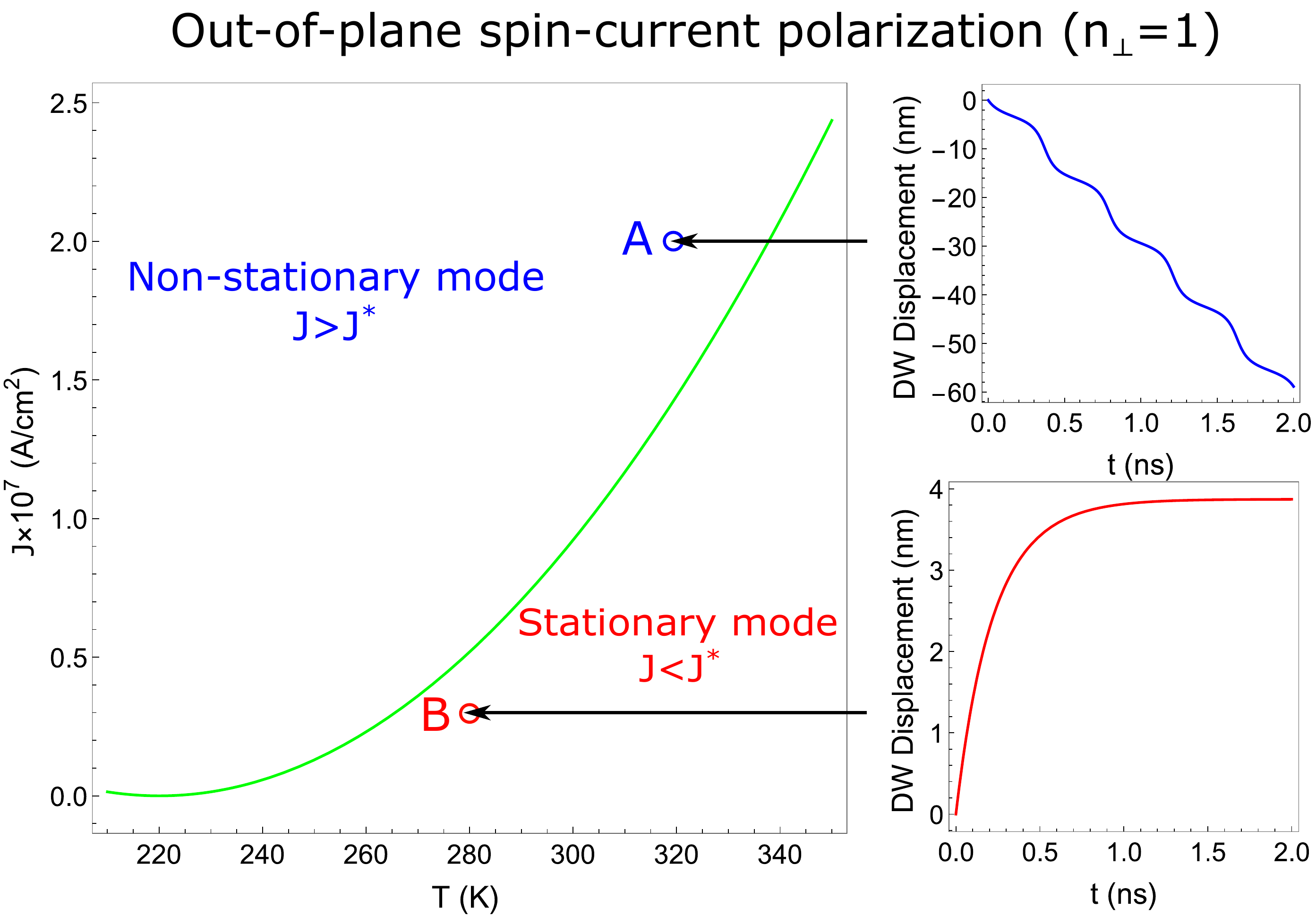}}
\caption{$J - T$ diagram which describes the ranges of a steady and non-steady motion of the DW, green curve shows the temperature dependence of the critical current $J^*$; the ranges above ($J > J^*$) and below ($J < J^*$) of the green curve correspond to non-stationary (post Walker) and stationary (Walker) mode of the DW, respectively. Insets show the time dependence of DW displacement in non-stationary range for point A ($T = 320$ K  and  $J = 2\cdot10^7$ A/cm$^2$ ) and stationary range for point B ($T = 280$ K and $J = 0.3\cdot10^7$ A/cm$^2$) of the diagram. All curves are plotted for the out-of-plane spin current polarization.}
\label{Fig3}
\end{center}
\end{figure}
\par
Let us analyse the DW dynamic in the presence of in-plane magnetic anisotropy $K_{\perp} \neq 0$ and $\textbf{n} = (0, 0, 1)$. We find out that there are two different regimes of the DW motion: steady ($\dot{\varphi} = 0$) and non-steady ($\dot{\varphi} \neq 0$). Let us discuss the non-steady one. An analytical solution to the system of differential equations (\ref{eq4}) can be written as:
\begin{equation}\label{Eq8}
     \tan\varphi = \dfrac{J^*}{J} +\sqrt{1 - \Big(\dfrac{J^*}{J}\Big)^2}\tan(\omega_0t - \varphi_0),
\end{equation}
where $J^* = \frac{4\pi \mathcal{M}^2\nu^2el}{ \hbar P_{DL}}$ is the critical current density, $\omega_0 =  \frac{\overline{\gamma}}{\overline{\alpha}}\frac{\pi \nu^2 \mathcal{M}\sqrt{1 -  (J^*/J)^2 }}{1 +(\nu/2\alpha_{\mathrm{eff}})^2}$, $\varphi_0 = \arctan\frac{J^*/J}{\sqrt{1 -  (J^*/J)^2 }}$. The non-stationary regime realises when the current density $J$ higher than critical current $J^*$ ($J > J^*$). This situation is represented on the $J - T$ diagram in Fig.~\ref{Fig3}, where the green curve is the temperature dependence of the critical current density $J^*$. The equation (\ref{Eq8}) describes the oscillation of the angle $\varphi$ in the non-stationary range of the $J-T$ diagram ($J > J^*$) and inset for the point A in Fig.~\ref{Fig3} shows the time dependence of the DW displacement in this range at fixed temperature $T = 320$ K and current dencity $J = 2\cdot10^7$ A/cm$^2$. Stationary regime of the DW  realises when the current density $J$ is lower than critical current $J^*$. Inset in Fig.~\ref{Fig3} for the point B ($T = 280$ K and $J = 0.3\cdot10^7$ A/cm$^2$) shows that after a small period of time $\sim 0.15$ ns the DW displacement stops changing in time. Therefore, the precession rate $\dot{\varphi} = 0$, velocity of the DW tends to zero and the magnetization freezes in the stable state, which corresponds to the equation $\sin2\varphi =  J/J^*$ as follows from the (\ref{eq3}). Hence stationary (Walker) mode in considered case corresponds to absence of DW motion, while non-stationary mode is responsible for DW motion.
\par
\begin{figure}[h!]
\begin{center}
\center{\includegraphics[width=0.9\linewidth]{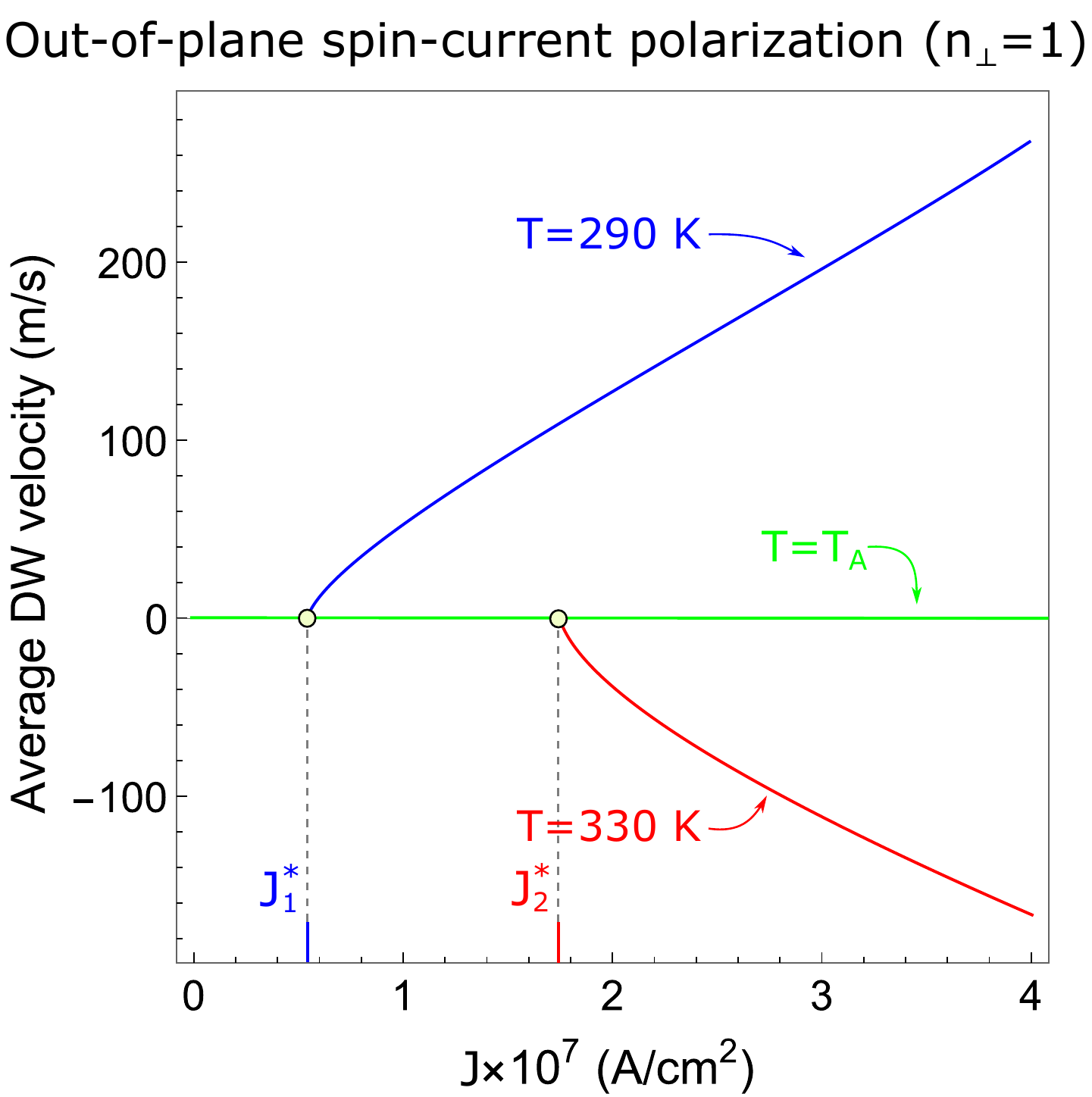}}
\caption{Average DW velocity in stationary and non-stationary mode as function of the electrical current density J; blue, green and red curves correspond to temperatures $T = 290$ K, $T = T_A$ and $T = 330$ K, respectively; $J^*_{1,2}$ are critical current densities which corresponds to $T = 290$ K and $T = 330$ K, respectively. All curves are plotted for the out-of-plane spin current polarization.}
\label{Fig4}
\end{center}
\end{figure}
\par
The dependence of the average DW velocity on the electrical current density in the stationary and non-stationary regimes is shown in Fig.~\ref{Fig4}. In the stationary mode the DW velocity is zero. Near the critical current $J^*$ an increase in the value of velocity occurs. However, in the non-stationary mode ($J > J^*$) the average DW velocity linearly increases. Note that in the angular momentum compensation point the average velocity is equal to zero (green curve in Fig.~\ref{Fig4}). Besides Fig.~\ref{Fig4} shows that velocity changes its sign passing through the T$_A$, which is demonstrated by blue ($T = 280$ K $<$ $T_A$) and red ($T = 330$ K $< T_A$) curves in Fig.~\ref{Fig4}. These results are consistent to the $J-T$ diagram in Fig. \ref{Fig3}.  It's important to note that in the non-stationary mode nonlinear spin waves can be excited and affect the dynamics of the DW. However, frequencies of the spin-wave in ferrimagnetic or antiferromagnetic materials lies in terahertz renge\cite{PhysRevB.96.100407, doi:10.1063/1.4958855, PhysRevLett.117.087203}. In contrast precession rate of quasi-antiferromagnetic vector lies in gigahertz range and we suppose that nonlinear spin waves have weak effect on the DW dynamics. Moreover, our model itself has a limitation (see Supplementary) in precession rate, which coincides with the frequencies of spin waves.    
\par
\begin{figure}[h!]
\begin{center}
\center{\includegraphics[width=0.9\linewidth]{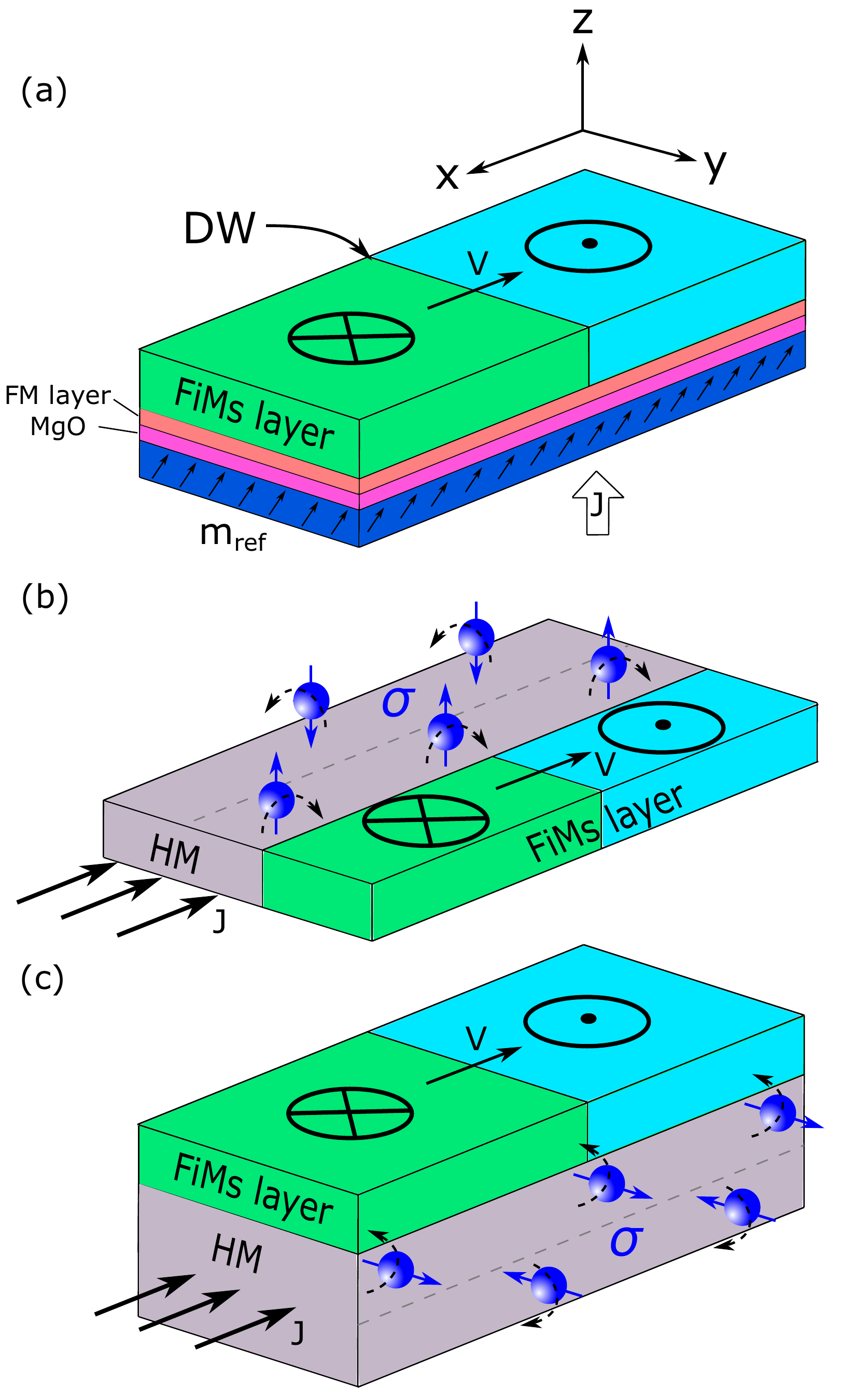}}
\caption{Examples of a) MTJ base, b)-c) spin Hall based structures, which can be used to observe reported DW motion and oscillation regimes in FIMs film or nanostripe. b) corresponds to perpendicular and c) - to planar direction of spin-current polarisation $\bm{\sigma}$.}
\label{Fig6}
\end{center}
\end{figure}
\par
Now, let us discuss the directions of the spin-current polarisation $\bm{\sigma}$ and type of torques, which can lead to effects mentioned above, and possibility of their experimental realization. As follows from the reported results, the damping (or anti-damping) spin transfer torque is responsible for considered motion and oscillation regimes for both planar and perpendicular spin-current polarisation $\bm{\sigma}$.
\par
The first possible way to create damping (or anti-damping) torque is to use magnetic tunnel junction (MTJ) structure. It is consist of free magnetic layer and polariser, which are separated by thin insulating material (usually MgO). In such a structure electric current flows perpendicularly to the plane and creates Slonczewski torque in the free layer, while spin-current polarisation $\bm{\sigma}$ direction is determined by magnetization direction of the polariser. Example of MTJ structure is presented in Fig.~\ref{Fig6}(a). The typical polarization value $P_{DL}$ in MTJ with ferromagnets is about 0.2-0.4. Hence one can add thin FM layer between MgO and FIMs, which is usually done even in classic MTJ to improve TMR and polarization values \cite{chanthbouala2011vertical}, to achieve the level of $P_{DL}=0.3$ used in our simulations.
\par
Another way to create damping (or anti-damping) torque is to use heavy metal / FIMs heterostructure. In such structure electric current flows through heavy metal (like Ta, W, Pt, Au etc.) in plane of the film and due to the spin Hall effect creates perpendicular spin current with polarization $\bm{\sigma}$, which is perpendicular to the both electric and spin currents. This spin current can create anti-damping torque in FIMs. The examples of spin Hall based geometry in case of perpendicular and planar polarization $\bm{\sigma}$ is presented in Fig.~\ref{Fig6}(b) and Fig.~\ref{Fig6}(c) respectively. The value of polarisation in these cases is equal to spin Hall angle. This angle can be up to 0.3 \cite{pai2014enhancement, RevModPhys.87.1213}, which again makes our $P_{DL}=0.3$ is reasonable. Moreover, it is possible to use topological insulator instead of heavy metal to achieve spin Hall angles more than 1 \cite{Mellnik-2014}, which significantly decrease required current densities.

\section{Discussion and Conclusion}

The theoretical study of the DW dynamics caused by the spin-current near the angular momentum compensation point is performed by using the Lagragian formalism. The non-linear dynamic equations describing the DW motion are derived from the effective Lagrangian of the two sublattice ferrimagnets. We analyse the DW motion at different directions of the spin-current polarizations and show the different types of magnetic heterostructures where this spin-current polarization can be realised. In the case of the out-of-plane polarizer ($n = (0, 0, 1)$) we analyse dependence of DW velocity and precession rate on temperature and current density. We foresee the possibility to generate oscillations of the quasi-antiferromagnetic vector $\textbf{L}$ with the frequencies by about 17 GHz at low current densities in the vicinity of the angular momentum compensation temperature. This oscillations are initiated by the damping (or anti-damping) component of the spin-transfer torque. This precession movement can be associated with a recent micromagnetic modelling of THz oscillation caused by a spin current in antiferromagnetic materials\cite{PhysRevB.99.024405} at high current densities. Furthermore, the DW velocity changes the direction passing trough this temperature and this effect is observed experimentally in GdFeCo ferrimagnet due to spin-current\cite{Okuno2019}. We explore the DW motion in the stationary (Walker) and non-stationary (post Walker) modes and construct the diagram that provides the values of current densities and temperatures for which these modes are realised. The model shows that in the Walker regime no DW motion occurs, while in the post Walker range DW velocity linearly increases with the current. Note, that the similar dependence of the DW velocity was observed due to the spin Hall effect\cite{PhysRevLett.121.057701} in the TbCo ferrimagnet sample in presence of external magnetic field. We also analyze the DW dynamics for the in-plane spin-current polarization and obtain the dependence of the DW velocity as a function of current in the Walker and post Walker regimes. Finally, we determine the directions of the spin-current polarisation $\bm{\sigma}$ and type of torques, which lead to effects mentioned above, and possibility of their experimental realization. These results may be useful for experimental studying of domain wall dynamics in ferrimagnets.  

This research has been supported by RSF grant No. 19-12-00432.

\providecommand{\noopsort}[1]{}\providecommand{\singleletter}[1]{#1}%

\end{document}